\documentclass[12pt,a4paper]{panl}
\usepackage{cite}
\usepackage{wrapfig}
\usepackage{graphicx}
\usepackage{amssymb}
\usepackage{amsfonts}
\usepackage{amsmath}
\usepackage{longtable}
\usepackage{rotating}
\usepackage{lscape}
\usepackage{epsfig}
\usepackage{multirow}
\usepackage{hyperref}

\originalTeX
\begin{document}

\title{Holographic equation of state matched with hadron gas equation as a tool for the study of the quark-gluon plasma evolution}
\maketitle
\authors{A.\,V. Anufriev$^{a}$\footnote{E-mail: anton.anufriev@spbu.ru},
V.N.\,Kovalenko$^{a}$\footnote{E-mail: v.kovalenko@spbu.ru}}
\setcounter{footnote}{0}
\from{$^{a}$\,Saint Petersburg State University}

\begin{abstract}
In this paper, we discuss the matching of the holographic equation of state with the equation of Hadron Resonance Gas for studying the nuclear matter properties within the framework of relativistic heavy-ion collisions. Machine learning methods are applied to the calibration of model's free parameters using the lattice QCD results for the physical values of quark masses. One of the most advanced procedures for matching is used with the function that approximate behavior of both models on particular limit adopted from NEOS equation. Final hadronic spectra are obtained within multi-staged numerical approach of the iEBE-MUSIC and SMASH-vHLLE packages. The code of relativistic hydrodynamics is modified by implementing a tabulated holographic equation of state, enabling simulations of quark-gluon plasma evolution with dynamically generated initial conditions via the 3D Monte Carlo Glauber Model and SMASH. Hybrid iSS+UrQMD and Hadron Sampler+SMASH approaches are utilized at the freeze-out stage.  \
\vspace{0.2cm}
\end{abstract}
\vspace*{6pt}


\label{sec:intro}
\section*{Introduction}

The study of the phase diagram of a quark-gluon plasma (QGP) is becoming especially relevant in modern research, since it is assumed that this special phase of matter has properties similar to relativistic liquid ~\cite{2005gfr}. In recent years, the relativistic hydrodynamics model ~\cite{2007du} has gained considerable popularity. The equation of state (EoS), whose general form remains unknown, is successfully approximated in the framework of lattice QCD at baryon potentials close to zero, assuming a smooth crossover transition in this region ~\cite{2014hot}. However, a `sign problem" arises due to the uncertainty of the fermionic determinant~\cite{2010Reeb} for non-zero values of $\mu_B$, which are usually associated with the presence of a critical point in the quark-gluon plasma (QGP). This allows us to consider more exotic options.

In 1998, Maldacena proposed a rigorous formulation of AdS/CFT invariance ~\cite{1997re}, which further allowed one to consider the advantages of this approach in the low-energy limit of string theory, opening the working domain of AdS/QCD duality~\cite{2005qh}. A striking example of this approach is the research cycle of I. Ya. Arefieva's theoretical group ~\cite{2017tdz}, which introduces an additional scalar dilaton field within the framework of the classical Einstein-Maxwell action by using the potential restoration method. The chiral limit in lattice quantum chromodynamics is effectively restored by selecting a deformation factor  with a certain set of free parameters for the initial ansatz, as was done, for example, in the reference ~\cite{2020byn}.

In our previous work ~\cite{2025AK1,2025AK2}, it was pointed out that holographic equations of state often remain theoretical and are rarely used to study available experimental data and numerical predictions. This allowed us to propose a practical way to implement the holographic approach in software packages for numerical simulation of the nuclear matter evolution in relativistic collisions of heavy ions. This study is devoted to the traditional problem of matching the equation of state in the framework of hydrodynamic modeling with the equation of a hadron gas at low temperatures. Having studied the experience of lattice equations on this issue, we have chosen one of the most advanced methods available at the moment. This allows us to eliminate problems with conservation laws that arose at earlier stages in the history of the development of such a matching ~\cite{2010hp}.

\section*{Approaches for calculations}

This study examines the equation of state obtained using the soft-wall bottom-up approach, which makes it possible to link the quasi-conformal theory of QCD with classical gravity in the AdS space of dimension 5, as proposed in ~\cite{2020byn}.

A special form of the initial ansatz is introduced.:

\begin{equation}\label{eq1}
ds^2=\frac{R^2}{z^2}B(z)\left[ -g(z)dt^2+dx^2+\left(\frac{z}{R}\right)^{2-\frac{2}{\nu}}dy_1^2+\left(\frac{z}{R}\right)^{2-\frac{2}{\nu}}dy_2^2+\frac{dz^2}{g(z)} \right],
\end{equation}
here, $R$ is the dimensional coefficient that would correspond to the AdS radius for the Poincare metric (we use $R=1$ for further calculations. This is a typical choice for theoretical work and does not reduce the generality of the above expressions), $g(z)$ is a blackening function that determines the thermodynamic behavior of a black brane. The $\nu$ parameter controls the spatial anisotropy of the metric components. In this case, $\nu=1$ corresponds to the isotropic case, and the choice of $\nu=4.5$, as shown in ~\cite{2016phb}, is the best for an anisotropic model in the context of restoring the dependence of the density on the multiplicity of charged hadrons experimentally observed on the collision energy.

The deforming factor $B(z)$ from equation \eqref{eq1} in the reference ~\cite{2024mmd} corresponds to the model of "light quarks" and was chosen in a way to restore the results of lattice QCD calculations in the range of $m_q\rightarrow 0$. Thus:

\begin{align} \label{eq2}
B(z)=\exp({2A(z)}), \\
A(z)=d \ln(az^2+1)+d \ln(bz^4+1). \label{eq3}
\end{align}

This approach allows one to obtain a relationship between the main thermodynamic quantities and the gravitational parameters of the black brane, the results of which were shown in our previous works.

After preparation of the resulting EoS for numerical simulation (will be described with more details in section 3), we implement it in the packages for relativistic hydrodynamics MUSIC~\cite{2010SJG} and VHLLE~\cite{2022SK}.

Final hadron spectra discussed in section 4 are obtained using a multi-stage approach that includes 3D MC Glauber\cite{2017bsr} and SMASH~\cite{2016zqf} models to find initial conditions for hydrodynamics. In addition, the iSS~\cite{2016zq} and Hadron Sampler~\cite{2015xea} packages are used at the freeze-out stage, and final hadron evolution is considered in the context of UrQMD~\cite{1998ca} and SMASH transport model in the "afterburner" mode.

Since hydrodynamic modeling requires detailed adjustment of the initial conditions and parameters of the model itself (for example, to include viscosity), we are just constructing a full methodology at this stage of our research and do not assume significant agreement of the predictions given in section 4 with experimental results. This task will be solved in later studies.

\section*{Setting up model parameters and matching a holographic equation with a hadron gas}

\begin{figure}[t]
\begin{center}
\includegraphics[scale=0.5]{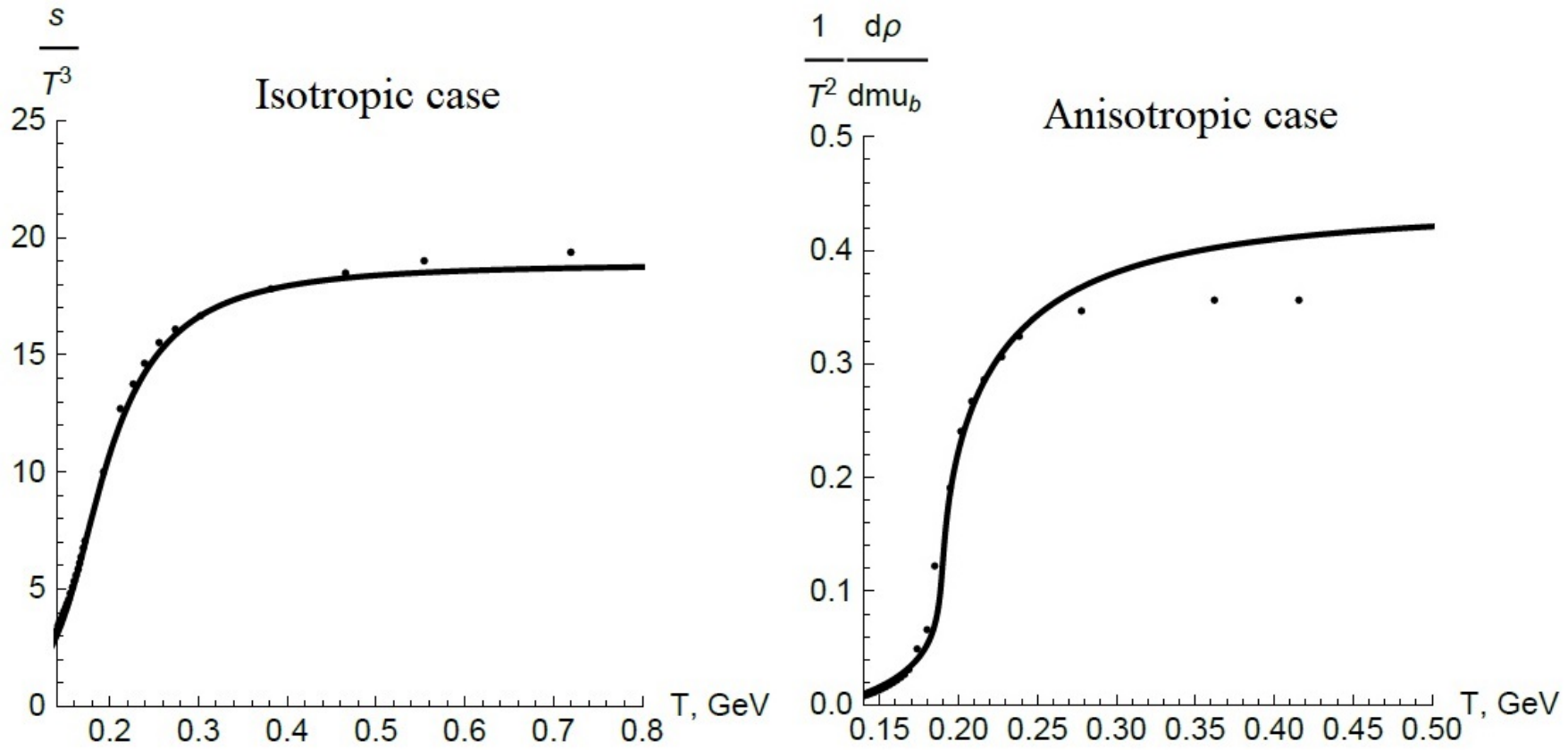}
\vspace{-3mm}
\caption{The result of fitting the ratio $\frac{s}{T^3}$ for isotropic (line in the picture on the left) and anisotropic (line in the picture on the right) models. The points for $T>0.18$ GeV are lattice results for the physical masses of quarks ~\cite{2007jq}. In the region of $T<0.18$GeV, we use as the points the results of the hadron gas equation obtained in the Thermal-FIST~\cite{2019pjl} package}
\end{center}
\vspace{-5mm}
\labelf{fig01}
\end{figure}

The main problem of this work is the issue of matching the equation of state used in numerical modeling with the hadron gas equation. This is widely discussed in the literature (see, for example, the review of ~\cite{2010hp}). Among the main arguments in favor of this procedure is the need to match the predictions of thermodynamics to the Cooper-Fry algorithm at the freeze-out ~\cite{74CF}, which implies a free hadron gas in the final state. In addition, this issue seems to be crucial for the multi-stage modeling used in this work, because Transport models also include the hadron gas equation as the main one.

Our previous research focused on a calibration of the hydrodynamic model on lattice results with experimental values of the masses of light mesons, which corresponds to the case of the physical masses of quarks. This is important since traditionally they use data obtained within the chiral or heavy limit for such tuning, which we do not consider as good approximations for realistic numerical modeling. At this stage, we consider the results of ~\cite{2007jq} as initial data, manually combining points for thermodynamic quantities we adjsut (their choice was discussed in our work ~\cite{2025AK1}) with predictions of the hadron gas for temperatures < 180 MeV (which roughly corresponds to thermodynamics after a phase transition). We obtain this set of points from the hadron gas equation within the Thermal FIST~\cite{2019pjl} package (which corresponds to the equation of state of SMASH and UrQMD). The results of the calibration using machine learning (the algorithm is given in ~\cite{2025AK2}) are shown in Fig. 1.

\begin{figure}[t]
\begin{center}
\includegraphics[scale=0.53]{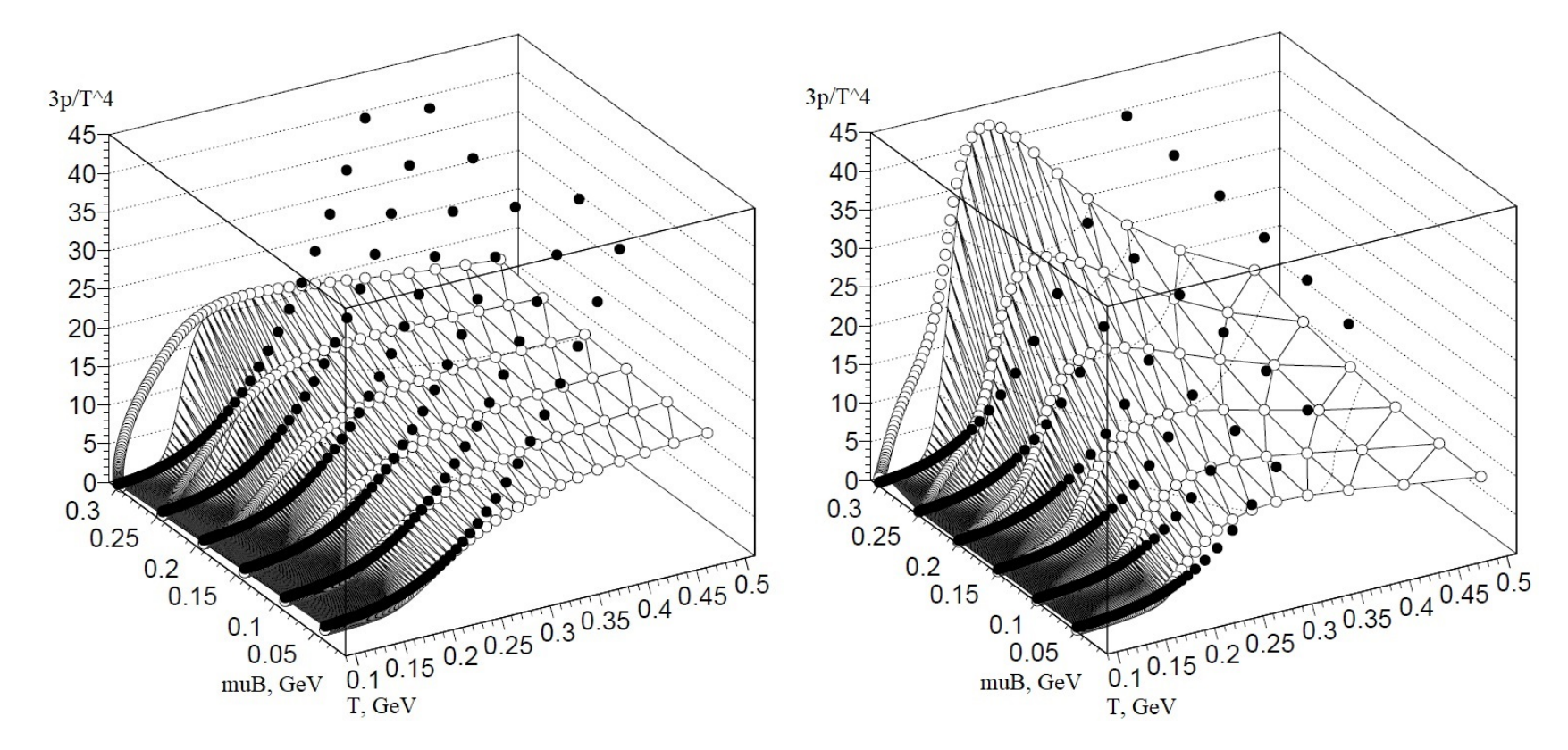}
\vspace{-3mm}
\caption{The results of matching the holographic equation with the hadron gas equation for the isotropic (left) and anisotropic (right) cases. The empty and shaded circles indicate the predictions of holography and hadron gas, respectively, and the black grid corresponds to the result with matching.}
\end{center}
\labelf{fig02}
\vspace{-5mm}
\end{figure}

This calibration method provides us with matching only at zero baryon potentials, whereas the main advantage of the holographic approach is the ability to study a wide area of the phase diagram. In order to incorporate these advantages, automatically ensuring compliance with conservation laws within the framework of hydrodynamics, we apply the approach of the authors of the NEOS equation, the details of which are described in detail in the work ~\cite{2019MSS}. The approach involves obtaining basic thermodynamic quantities in the form of a linear combination of holographical approach and the hadron gas with dynamic coefficients depending on temperature and the baryonic potential. This dependence is provided by the choice of a `switching function", is taken in our work, as in the above-mentioned study, in the following form:

\begin{equation*}
f(T,\mu_b)=\tanh \left(T-\frac{ {T_c(\mu_b)}}{{\Delta T_c}}\right)
\end{equation*}

The matching temperature $T_c$ is obtained from a detailed study of chemical freeze-out, and $\Delta T_c=0.1T_c(0)$.

This approach allows for a smooth approximation for both approaches in the corresponding temperature ranges, as shown in Fig. 2.

\section*{Results and conclusions}

In Figs. 3 and 4 we present the results of multi-stage modeling within the iEBE-MUSIC and SMASH-vHLLE using a holographic equation of state matched with the hadron gas equation in the kinematic region of the NA49 experiment for the most central collisions at $\sqrt{s} = 8.9$ GeV ~\cite{2002pzu}.

\begin{figure}[t]
\begin{center}
\includegraphics[scale=0.5]{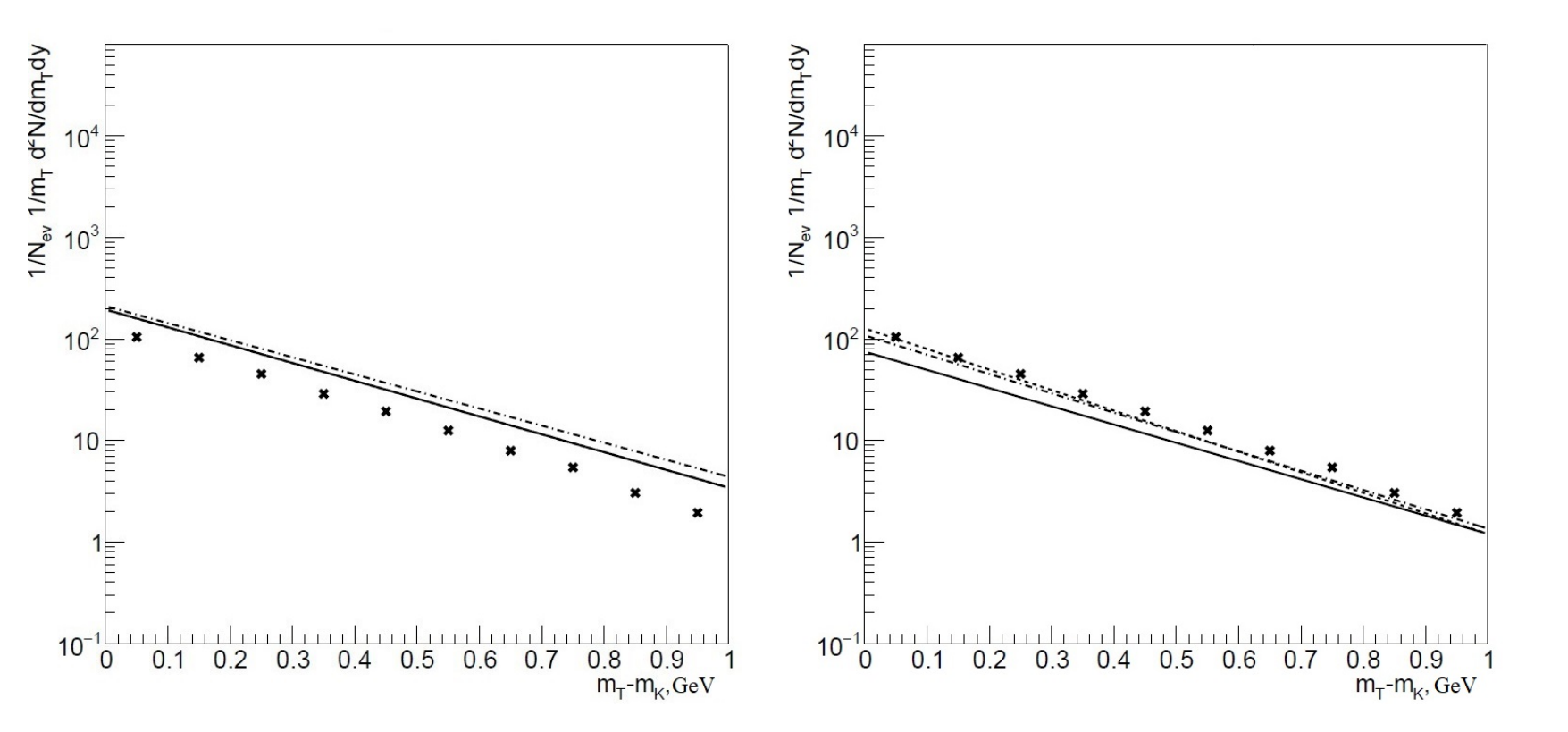}
\vspace{-3mm}
\caption{The results of the $m_T$ spectra calculations for $K^{+}$ in the MUSIC-UrQMD (left) and SMASH-vHLLE (right) approaches using the equation of state with matching. The solid line corresponds to the isotropic equation of state, when the dotted line is the anisotropic equation. The dot-and-dash line corresponds to the reference equation (NEOS ~\cite{2019MSS} in the case of iEBE-MUSIC, AZ Hydro~\cite{2012HFR} for SMASH-vHLLE). The points were taken for experimental data NA49 ~\cite{2002pzu}}
\end{center}
\labelf{fig03}
\vspace{-5mm}
\end{figure}

Due to the high sensitivity of hydrodynamics to initial conditions, it is important to point out that in the two cases given, the same initial parameters for modeling were set for 3D MC Glauber and SMASH with all three equations used to ensure a clean comparison of the results obtained.

It was impossible to expect significant differences in the results obtained in the comparison with the predictions from our previous work (~\cite{2025AK2}, Figures 7 and 8), since the changes in the approach are not significant compared, for example, with the alternative choice of the deforming factor \eqref{eq3}. However, by adapting the matching procedure from the NEOS equation, we should expect an improvement in the agreement of our results with the hydrodynamic predictions for this reference, which is indeed observed in comparison with previous results.

\section*{Acknowledgements}
The authors acknowledge Saint-Petersburg State University for a research project 103821868.

\section*{Conflict of interest}

The authors of this paper declare that they have no conflict of interest.

\end{document}